\documentclass[aps,prl,twocolumn,showpacs,10pt]{revtex4-1}
\usepackage{amssymb}
\usepackage{amsmath}
\usepackage[pdftex]{graphicx}
\usepackage{dsfont}
\usepackage{bm}
\usepackage[normalem]{ulem}
\usepackage{hyperref}
\hypersetup{
    colorlinks,%
    citecolor=blue,%
    filecolor=blue,%
    linkcolor=blue,%
    urlcolor=blue
}

\newcommand{\mueV}{~\ensuremath{\mu\text{eV}}}
\providecommand{\ket}[1]{\vert #1\rangle} 
\providecommand{\mean}[1]{\langle #1 \rangle} 

\usepackage{color}

\begin{document}
\title{Helical States  in Curved Bilayer Graphene}
\author{Jelena Klinovaja}
\author{Gerson J.~Ferreira}
\author{Daniel Loss}
\affiliation{Department of Physics, University of Basel, Klingelbergstrasse 82, CH-4056 Basel, Switzerland}
\date{\today}

\begin{abstract}
We study spin effects of  quantum wires formed in  bilayer graphene  by electrostatic confinement.
With a proper choice of the confinement direction, we show that in the presence of  magnetic field, spin-orbit interaction induced by curvature, and intervalley scattering,  bound states
emerge that are helical.  The localization length  of these helical states can be modulated by the gate voltage
 which enables the control of the tunnel coupling between two parallel wires.
Allowing for proximity effect via an $s$-wave superconductor, we show that the helical modes
give rise to Majorana fermions in bilayer graphene.
\end{abstract}

\pacs{73.22.Pr,  75.70.Tj, 73.63.Fg, 72.25.-b}
\maketitle

\paragraph{Introduction.} Graphene and its derivatives \cite{novoselov2005two,CastroNeto2009Review,dresselhaus_book,arxiv_review_McCann}, such as bilayer graphene (BLG) and carbon nanotubes (CNT), have attracted wide interest due to its peculiar bandstructure with low energy excitations described by Dirac-like Hamiltonians. Moreover, these materials are usually placed on substrates, which allows high control of its geometry, doping, and placement of metallic gates \cite{Geim2007Rise, bilayer_yacoby_2010, graphene_marcus_2011, bilayer_yacoby_2012, arxiv_Vandersypen}.
Topological insulators  were  predicted for graphene \cite{Kane2005QSHE}, but later it was found that the intrinsic spin-orbit interaction (SOI) is too weak \cite{min_soi_2006, gmitra_spin_orbit_d_orbitals_2009}.
For BLG, first-principle calculations also show weak SOI  \cite{Konschuh2012SOBilayer, Mireles_SOI_2012}.
In an other proposal, topologically confined bound states were predicted to occur in BLG where a gap and
band inversion is enforced by gates~\cite{martin2008topological}.
Quite remarkably, these states are localized in the  region where the voltage changes sign, are independent of the edges of the sample, and propagate along the direction of the gates,
thus forming effectively a quantum wire~\cite{martin2008topological, qiao2011electronic, Zarenia2011ChiralGap}.
At any fixed energy, the  spectrum inside the gap
is topologically equivalent to four Dirac cones, each cone consisting of a pair of states with opposite momenta.

\begin{figure}[t]
 \centering
 \includegraphics[width=\columnwidth,keepaspectratio=true]{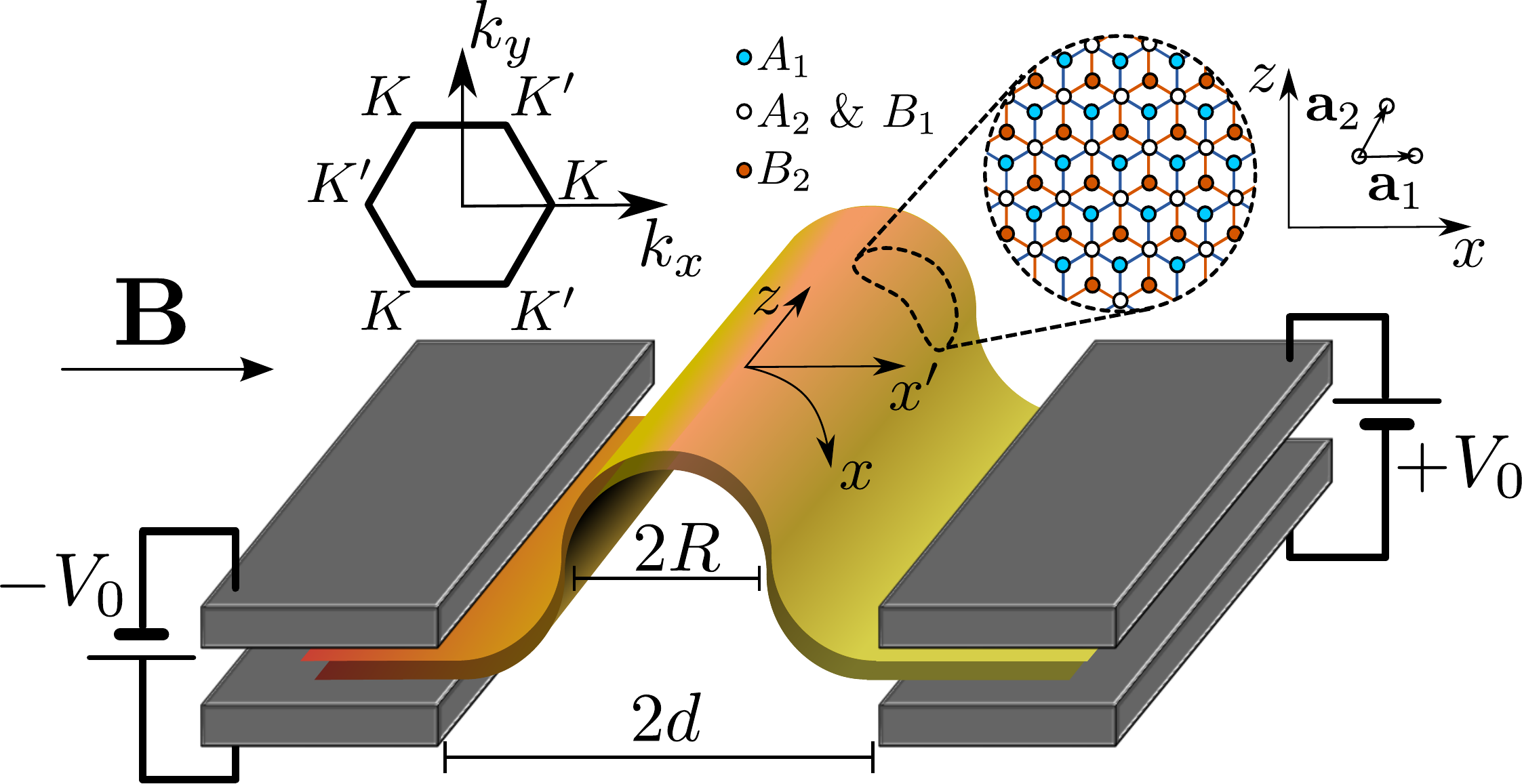}\\
 \caption{A  bilayer graphene (BLG) sheet with a fold at $x=0$ along the $z$-axis is placed between two pairs of gates that are set to opposite polarities $\pm V_0/2$,
 inducing the bulk gap.
 There are mid-gap bound states,  localized in transverse $x$-direction around $x=0$. At the same time, they freely propagate along the $z$-direction,
 forming an effective quantum wire~\cite{martin2008topological}.
 An externally applied magnetic field $\mathbf{B} = B {\bf e}_{x'}$ breaks time-reversal symmetry.
The spin-orbit interaction $\beta$ is induced by the curvature of the wire, which is characterized by the radius $R$. In the insets we show the BLG structure in momentum (left) and real (right) space for a chosen chirality $\theta=0$. The edges of the BLG sheet can be arbitrary.
}
 \label{fig:system}
\end{figure}

\begin{figure}[ht!]
 \centering
 \includegraphics[width=\columnwidth,keepaspectratio=true]{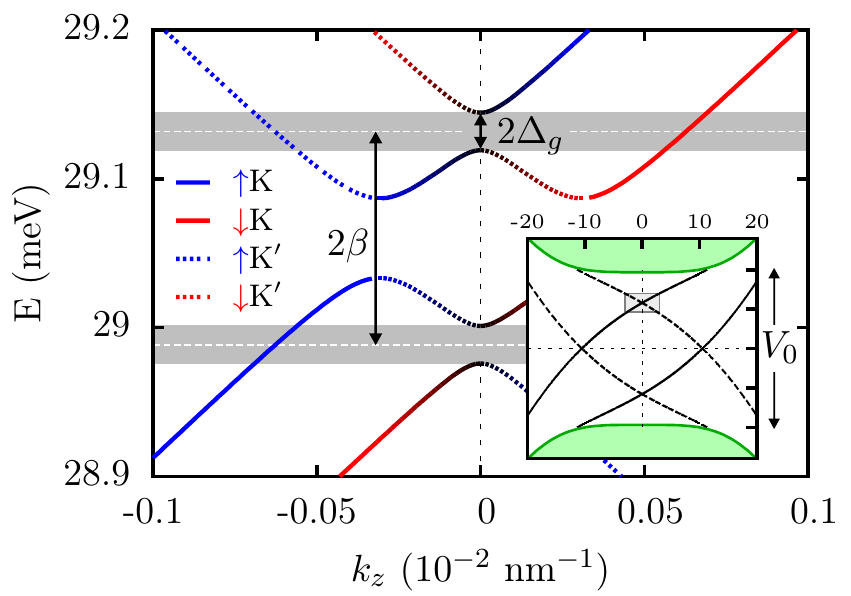}
  \caption{The  spectrum of the BLG structure for $V_0/2 = 50$~meV and chirality $\theta=0$.
The green area in the inset corresponds to the bulk spectrum. The mid-gap bound states for valleys $K$ (full line) and $K^\prime$ (dashed line) have  opposite velocities.
The main figure shows the details of $K$-$K^\prime$ crossing region (shaded region in the inset). The curvature induced SOI
shifts spin-up and spin-down levels in opposite directions by the SOI parameter $\beta$. A magnetic field $\mathbf{B}$ assisted by  intervalley scattering $\Delta_{KK^\prime}$ results in the anti-crossing gap $2\Delta_g$ of two Kramers partners at $k_z=0$. If the chemical potential $\mu$ is tuned inside the gap  [shaded area with $\mu\approx (V_0/2\sqrt{2})\pm\beta$], the system  is equivalent to three Dirac cones (only one is shown in the main Figure), resulting in the helical mode regime.
Here,  $\beta\approx60$\mueV\ ($R=5\ \rm nm$),  $\Delta_{KK^\prime} = 30$\mueV, and $\Delta_Z = 30$\mueV, so the opened gap is $2\Delta_g \approx 30$\mueV$\approx 300$~mK.
}
 \label{fig:bandstructureV100}
\end{figure}

The spin degrees of freedom in such BLG wires, however, have not been addressed yet. It is the goal of  this work to include them
and to show that they give rise to striking effects.
 In particular, we uncover a mechanism enabling {\it helical modes} propagating along the wires.
 In analogy to  Rashba nanowires~\cite{Streda}, topological insulators~\cite{Kane_Hassan_RMP}, and CNTs~\cite{klinovaja2011helical, klinovaja_MF_2012},
 such modes provide the platform
 for a number of interesting effects such as spin-filtering and Majorana fermions~\cite{Alicea_MF_2012}.
 Here, the SOI plays a critical role, and in order to  substantially
 enhance it, we consider a BLG sheet with local curvature as shown in Fig.~\ref{fig:system}.
 Two pairs of top and bottom gates define the direction of the quantum wire which is chosen in such a way that it corresponds to a `semi-CNT' of zigzag type.
 In this geometry, the energy levels of the mid-gap states cross in the center of the  Brillouin zone. A magnetic field
 transverse to the wire in combination with intervalley scattering
 leads to an opening of a gap, $2\Delta_g$, between two Kramers partners at zero momentum, see Fig.~\ref{fig:bandstructureV100}.
 As a result, the number of Dirac cones changes from even
  (four) to odd (three), and the wire becomes helical with opposite spins being transported into opposite directions.
  In the following we derive the spectrum and its characteristics analytically and confirm these results by independent numerics. We also address the physics of Majorana fermions which emerge when the wire is in proximity contact to an $s$-wave superconductor.

\paragraph{Curved  bilayer graphene with SOI.}
We consider a gated curved  bilayer graphene
with a magnetic field $\bf B$ (along the $x'$-axis) applied perpendicular to the direction of the fold (along the $z$-axis), see Fig. \ref{fig:system}. We begin with a description of the bilayer graphene in the framework of the tight-binding model \cite{dresselhaus_book,arxiv_review_McCann}. Each layer is a honeycomb lattice composed of two types of non-equivalent atoms $A_1$ ($A_2$) and $B_1$ ($B_2$) and defined by two lattice vectors ${\bf{a}}_1$ and ${\bf{a}}_2$. We focus here on AB stacked bilayer, in which two layers are coupled only via atoms $A_2$ and $B_1$ (see Fig.~\ref{fig:system}) with a hopping matrix element $t_\perp$ ($t_\perp\approx 0.34\ \rm{eV} $). By analogy with CNTs~\cite{dresselhaus_book}, we introduce  a chiral angle $\theta$ as the angle between ${\bf{a}}_1$ and  the $x$-axis.

The low-enegy physics is determined by two valleys  defined as ${\bf K}=-{\bf K}^\prime=(4\pi/3a) (\cos \theta,\ \sin \theta)$, where $a=|{\bf a}_1|$.
The corresponding Hamiltonian in momentum space is written as
\begin{equation}
H_0=\hbar \upsilon_F (k_x \sigma_1 + \tau_3 k_z \sigma_2) + \frac{t_\perp}{2}(\sigma_1 \eta_1+\sigma_2\eta_2)-V\eta_3,
\end{equation}
where the Pauli matrices $\sigma_i$ ($\eta_i$) act in the sublattice (layer) space, and the Pauli matrices $\tau_i$ act in the valley space. Here, $\upsilon_F=\sqrt{3}ta/2\hbar$ is the Fermi velocity ($\upsilon_F\approx 10^8\ \rm{cm/s}$), with $t\approx 2.7\, \rm{eV}$ being the intralayer hopping matrix element.
 The $k_x$  ($k_z$) is the transversal (longitudinal) momentum calculated from the  points ${\bf K}$ and ${\bf K}^\prime$. The potential difference between the layers opens up a gap $2|V|$ in the bulk spectrum, while a spatial modulation,
i.e. $V \rightarrow V(x)$,
breaks the translation invariance along the $x$-direction, thus only the total longitudinal momentum ${\bf{K}}^{(\prime)}_z+k_z$ remains a good quantum number.

The Hamiltonain $H_0$ can be simplified for small voltages, $|V|\ll t_\perp$, by integrating out the $A_2$ and $B_1$ degrees of freedom, which correspond to much higher energies $E\approx t_\perp$. The effective Hamiltonian becomes
\begin{equation}
\tilde{H}_0 = -V\gamma_3 - \frac{\hbar^2 \upsilon_F^2}{t_\perp } \left( k_x^2-k_z^2\right)\gamma_1-\frac{2\hbar^2 \upsilon_F^2}{t_\perp} k_x k_z \tau_3 \gamma_2,
\end{equation}
where the Pauli matrices $\gamma_i$ act in the space of $A_1$ and $B_2$ atoms.
 If the  voltage changes sign at $x=0$ [for example, $V(x)=-V(-x)$],
this results in the closing and reopening of the gap.
As a consequence, bound states, localized around  $x=0$, emerge within the bulk gap~\cite{martin2008topological}.
The eigenstates of $\tilde{H}_0$ are characterized by $k_z$ and the valley degree of freedom $\tau=\pm 1$. For a step-like kink potential \mbox{$V(x)=(V_0/2)\ \text{sgn}(x)$} the energy spectrum is shown
in the inset of Fig.~\ref{fig:bandstructureV100}.

Now we include spin and
 aim at the realization of helical modes in BLG, which requires an analysis of the spin-full mid-gap states.
At any fixed energy in the bulk gap, there are $2\times 4$ states, where the factor $2$ arises from spin-degeneracy. This means that the spectrum is topologically equivalent to four Dirac cones, each cone  consisting of a pair of states with opposite momenta.
 On the other hand, helical modes are typical for systems with an odd number of Dirac cones. To effectively eliminate one Dirac cone at given chemical potential, the spin-degeneracy should be lifted by a magnetic field $B$, giving rise to  a new gap.
Obviously, the opening of such a gap is possible only if there is level crossing in the system. The spectrum of the mid-gap states has support around $K$ and $K^\prime$. Therefore, if
these points, projected onto the $k_z$-axis, are separated from each other, no crossing can occur. We thus see that the chiral angle $\theta$ is of a crucial importance for our purpose and the optimal choice is $\theta=0$ (or very close to it).
In this case, ${\bf K}_z={\bf K}^\prime_z=0$, and the level crossing occurs in the center of the Brillouin zone, at $k_z=0$, see  inset of Fig.~\ref{fig:bandstructureV100}.
We emphasize that in contrast to  nanoribbons~\cite{qiao2011electronic}  the form of the edges of the BLG sheet does not matter provided the distance between edges and wire-axis is much larger than the localization length $\xi$ of the bound state.

Next, we allow also for spin-orbit interaction in our model. While the intrinsic SOI is known to be weak for graphene~\cite{min_soi_2006, gmitra_spin_orbit_d_orbitals_2009}, the strength of SOI in CNT is enhanced by curvature \cite{kuemmeth_exp_cnt_2008, izumida_soi_cnt_2009, klinovaja2011helical, klinovaja2011carbon}. To take advantage of this enhancement, we consider a folded BLG which is analogous to a zigzag semi-CNT with $\theta=0$. All SOI terms that can be generated in second-order perturbation theory are listed in Table I of Refs.
\cite{klinovaja2011helical,klinovaja2011carbon}. From  these terms only $H_{so}= \beta\tau_3 s_z$ is relevant for our problem; first, it is the largest term by magnitude, and second, it is the only term which acts directly in the $A_1$-$B_2$ space. Here, $s_i$ is the Pauli matrix acting on the electron spin, and $i=x,y,z$. The value of the effective SOI strength $\beta$ depends on the curvature, defined by the radius $R$, and is given by $\beta\approx0.31\ \rm{meV}/R\rm{[nm]}$~\cite{klinovaja2011helical}.

In the  presence of SOI, the states can still be  characterized by the momentum  $k_z$, valley index $\tau=\pm1$, and spin projection $s=\pm 1$ on the $z$-axis.
The spectrum of  $\tilde{H}_0 + H_{so}$ can be obtained from the one of $\tilde{H}_0$ by simply shifting $E\rightarrow E- \beta \tau s$.  This transformation goes through the calculation
straightforwardly,
and the spectrum in the presence of the SOI becomes
\begin{align}
E =\beta \tau s\pm  \left(  \frac{\hbar \upsilon_F k_z \tau}{2\sqrt{t_\perp}} \pm \sqrt{\frac{(\hbar \upsilon_F k_z)^2}{4 t_\perp}+\frac{V_0}{2\sqrt{2}}}\right)^2 \mp \frac{V_0}{\sqrt{2}}.
\end{align}
The spin degeneracy is lifted by the SOI, giving a splitting $2\beta$. As shown in Fig.~\ref{fig:bandstructureV100}, the level crossings occur between two Kramers partners at $k_z=0$: $\ket{K, \uparrow}$ crosses with $\ket{K^\prime, \downarrow}$, and $\ket{K, \downarrow}$ crosses with $\ket{K^\prime, \uparrow}$.
The $KK^\prime$-crossing can occur provided $|\theta|< \sqrt{3(1+\sqrt{2})t_{\perp}V_0}/4\pi t $. For the values from Fig.~\ref{fig:bandstructureV100}, we estimate this bound to be about $1^{\circ}$.
As mentioned before, to open a gap at $k_z = 0$, one needs first a magnetic field perpendicular to the SOI axis to mix the spin states, and second a $K$-$K^\prime$ scattering to mix the two valleys.
Such valley scattering  is described by the Hamiltonian $H_{sc}=\Delta_{KK^\prime}^s\tau_1+\Delta_{KK^\prime}^a\tau_1\gamma_3$, where $\Delta_{KK^\prime}^s+\Delta_{KK^\prime}^a$  ($\Delta_{KK^\prime}^s-\Delta_{KK^\prime}^a$) is the scattering parameter for the bottom (top) layer of the BLG. The Zeeman Hamiltonian for a magnetic field $\bf B$ applied along the $x'$-direction is given by $H_Z=\Delta_Z s_{x}$, with $\Delta_Z=g^*\mu_B B/2$, where $\mu_B$ the Bohr magneton. Here,
$g^*$ is an effective $g$-factor due to the curvature of the fold and the localization of the bound state. Since $s_{x'}=s_x \cos\varphi + s_y\sin\varphi $ depends on $x$
via  the azimuthal angle $\varphi (x)$ of the fold,
we replace $s_{x'}$ by an average over the orbital part of the bound state wave function.
This  results in $2/\pi<g^*/g<1$, the precise value being dependent on the localization length, where $g$ is the bare $g$-factor of graphene.

Using second order perturbation theory for $\beta>\Delta_{KK^\prime}^s$, $\Delta_Z$,
we find that the  gap opened at $k_z=0$ is given by
\begin{equation}
 \Delta_g = \frac{\Delta_{KK^\prime}^s\Delta_Z}{\beta},
\end{equation}
see Fig.~\ref{fig:bandstructureV100}, which also contains numerical estimates for realistic parameters.
We note that $\Delta_g$ is enhanced by electron-electron interactions \cite{braunecker_wire_2010}, however, we neglect this supportive effect herein.
If the chemical potential is tuned inside the gap $2\Delta_g $ [$\mu\approx(V_0/2\sqrt{2})\pm \beta$], there are three right- and three left-propagating modes.
Four states at finite momentum (two left-moving and two right-moving states) are only slightly affected by the magnetic field and thus can still be considered to carry opposite spins, meaning that the total spin transfer is close to zero and these modes are not contributing to spin-filtering. In contrast to that, the two modes with $k_z\approx 0$ are {\it helical modes}: they have opposite velocities and opposite spins.
Thus, similar to  Rashba nanowires~\cite{Streda}, the BLG quantum wire can be used as a spin filter device.

\begin{figure}[t]
 \centering
 \includegraphics[width=\columnwidth,keepaspectratio=true]{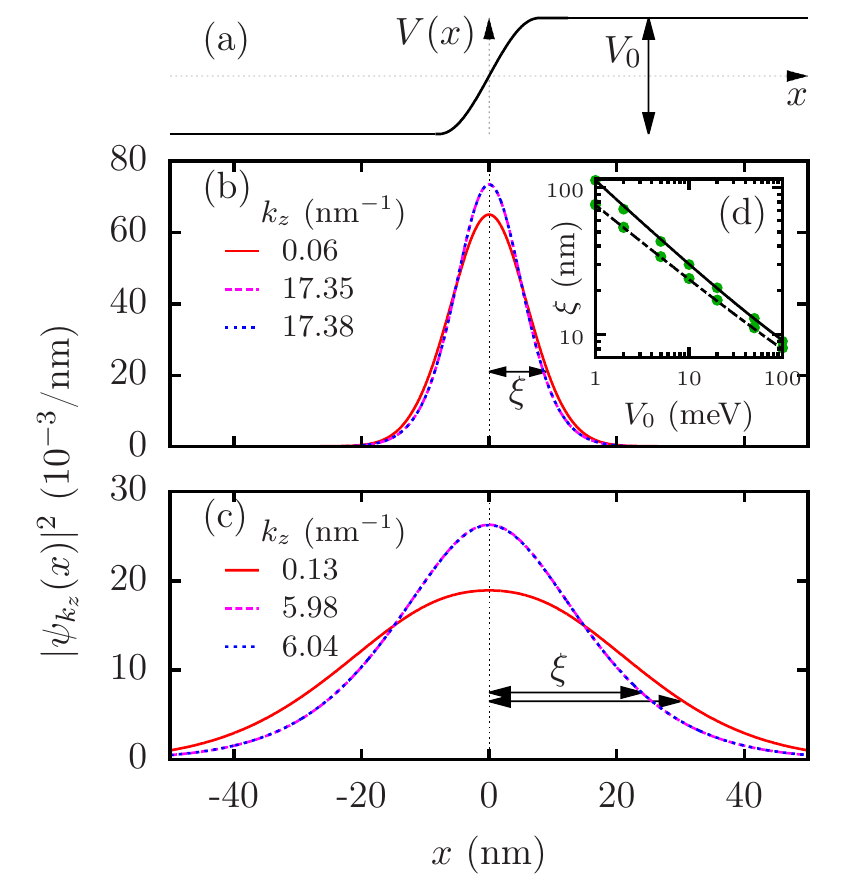}
 \caption{(a) The profile of the gate potential $V(x)$ along the curved BLG. The reversed polarity at the two ends gives rise to mid-gap states,
 localized in $x$-direction. The density profile of three right-moving states, whose energy $E$ is inside the gap $\Delta_g$, allows us to estimate the localization lengths $\xi$: (b)
$V_0 = 100$~meV, $E = 29.13$~meV, $d=R$ (dashed white line in Fig.~\ref{fig:bandstructureV100}), $\xi \approx 9$ and $8$~nm,  and (c) $V_0 = 10$~meV, $E = 3.49$~meV, $d=R$, $\xi \approx 30$ and $24$~nm. We note that states with larger momenta have shorter localization lengths.
(d) The localization length follows approximately $\tilde{\xi} = \xi - \xi_0 \propto 1/\sqrt{V_0}$. The circles are extracted from our numerical calculations, $\xi = \mean{x^2}$, for energies in the middle of the gap $\Delta_g$, equivalent to the white dashed line in Fig.~\ref{fig:bandstructureV100}. The lines are fits, $\tilde{\xi} \propto 1/V_0^p$, for
the states at large $k_z$ ($p=0.59$), shown as dashed lines,
and for the states near $k_z = 0$ ($p=0.52$), shown as full lines.}
 \label{fig:DensAndDecay}
\end{figure}

Moreover, if the BLG is brought into proximity to an $s$-wave superconductor,
the states with opposite momenta and spins get paired. Working in the linearized model of left-right movers~\cite{Lin_Model}, we obtain  the effective Bogoliubov-de Gennes Hamiltonian for each of the three pairs, $j=1,2,3$, written in Nambu space,
\begin{equation}
H_s^j = \hbar \upsilon_j k_j \chi_3+\Delta_s \omega_2 \chi_2,
\end{equation}
where $\upsilon_j$ is the velocity for the $j$th pair at the Fermi level and $\Delta_s$ is  the strength of the proximity-induced superconductivity, and the Pauli matrices $\chi_i$ ($\omega_i$) act in the left-right mover (electron-hole) space. We note that we are in the regime corresponding to strong SOI where we keep only the slowest decaying contributions of the
wave functions~\cite{Lin_Model}.
To determine the potential existence of MFs in the system, one can study the topological class of  $H_s^j$ \cite{Topol_classification}. This Hamiltonian  belongs to the topological class BDI. However, by analogy with multi-band nanowires \cite{Tewari}, additional scattering between states would bring the system into the D class. An alternative way of classification, which determines explicitly the number of MF bound states, is to study the null-space of the Wronskian associated with the Schr\"odinger equation~\cite{Wronskian_2012}. In our case, we find three MFs at each wire end in the topological phase defined by $\Delta_g ^2\geq \Delta_s^2+{\delta \mu}^2$, where
${\delta \mu}$ is the chemical potential counted from the mid-gap level $\Delta_g$.
These MFs are generically hybridized into one MF and one non-zero energy fermion by perturbations such as  electron-electron interactions and interband scattering.

\paragraph{Numerical calculation.}
Above we have studied the system analytically, assuming a step-like potential. In this section we compare our results with the numerical solution of the
Schr\"odinger equation for the effective Hamiltonian $\widetilde{H}_0+H_{so}+H_{sc}+H_Z$, with a  more realistic (smooth) potential, $V(x) = (V_0/2) \tanh(x/d)$, where $d$ is the distance between the gates.
The spin-orbit interaction $\beta(x)$ is finite only within the curved region of the BLG sheet.
Along the $z$-direction, the system is translationally invariant, so the envelope function is given by $\Psi(x,z) = e^{ik_zz} \psi_{k_z}(x)$. The profile of $\psi_{k_z}(x)$ is presented in Fig. \ref{fig:DensAndDecay}.
The localization length follows a power law $\xi-\xi_0 \propto 1/V_0^p$, with $p \approx 1/2$, and the shift $\xi_0 < d$ is due to the finite distance between gates. In the limit $d \rightarrow 0$, where the analytical solution is applicable, the localization length is essentially given by $\xi = 2^{5/4} \hbar \upsilon_F /\sqrt{V_0 t_\perp}$~\cite{martin2008topological}, since corrections due to SOI are of negligible higher order in $\beta$.
\paragraph{Tunnel junction.} The dependence of $\xi$ on the potential $V_0$ can be exploited to couple parallel wires.
For instance, consider two similar quantum wires,  running parallel to each other at a distance $D$.
If $\xi\ll D$ for each wire, then they are completely decoupled. However,
 lowering the potential in both wires {\it locally} around a point  $z_0$ on the $z$-axis, such that  $\xi_0 \approx D$, we can enforce wavefunction overlap, leading to a transverse tunnel junction between the two wires at $z_0$. In this way, an entire network of helical wires can be envisaged.
We mention that such networks could provide a platform for implementing braiding schemes for MFs~\cite{Alicea_braiding}.

\paragraph{Conclusions.} The confinement of states in BLG into an effective quantum wire is achieved
by  pairs of gates with opposite polarities, leading to eight propagating modes \cite{martin2008topological}. If the direction of the wire is chosen such that the chiral angle vanishes,
both valleys $K$ and $K^\prime$ are projected onto zero momentum $k_z$. The SOI, substantially enhanced by curvature, defines a spin quantization axis and splits spin-up and spin-down states. A magnetic field
assisted by intervalley scattering opens up a gap at the center of the Brillouin zone. If the chemical potential is tuned inside the gap, three right- and three left-propagating modes emerge, so that the system possesses helical modes, which are of potential use for spin-filtering. In the proximity to an $s$-wave superconductor, the BLG wire hosts Majorana fermions arising from the helical modes.
By  locally changing the confinement potential and thus the localization lengths, parallel wires can be tunnel coupled. This mechanism can be used to implement braiding
of MFs in bilayer graphene.
\acknowledgments
This work is supported by the Swiss NSF, NCCR Nanoscience, and NCCR QSIT.


\end{document}